\documentclass{webofc}
\usepackage{cancel}
\usepackage[utf8]{inputenc}
\usepackage[T1]{fontenc}
\usepackage{eurosym,rotating,multirow,overpic}
\usepackage{amsmath,amssymb,amstext}
\usepackage{graphicx}
\usepackage{units}
\usepackage{color,xcolor}

\usepackage[absolute,overlay]{textpos}
\newcommand{\beq}{\begin{equation}}
\newcommand{\eeq}{\end{equation}}
\newcommand{\ba}{\begin{eqnarray}}
\newcommand{\ea}{\nonumber \end{eqnarray}}
\newcommand{\be}{\begin{eqnarray}}
\newcommand{\ee}{\nonumber \end{eqnarray}}

\newcommand{\bs}{\begin{slide}}
\newcommand{\es}{\end{slide}}

\newcommand{\er}{$\pm$}

\newcommand{\bc}{\begin{center}}
\newcommand{\ec}{\end{center}}
\newcommand{\bi}{\begin{enumerate}}
\newcommand{\ei}{\end{enumerate}}

\begin{document}
\title{Glueballs in Radiative $J/\psi$ Decays}
%
%

\author{\firstname{Eberhard} \lastname{Klempt}\inst{1}\fnsep\thanks{\email{Klempt@hiskp.uni-bonn.de}}
}

\institute{HISKP der Rheinischen Friedrich-Wilhelms-Universit\"at,
Nu\ss allee 14-16, 53115 Bonn, Germany
          }

\abstract{The scalar glueball is observed in
 a coupled-channel analysis of
the $S$-wave amplitude from BESIII data on radiative $J/\psi$ decays
and further data. Ten scalar isoscalar resonances were required to
fit the data. Five of them were interpreted as mainly-singlet, five
as mainly-octet resonances in SU(3). The yield of resonances showed
a striking peak with properties expected from a scalar glueball:
\begin{itemize}
\item $G_0(1865)$ is produced abundantly in radiative $J/\psi$ decays above a very
low background. Its mass is $1\sigma$ compatible with the mass
calculated in unquenched lattice QCD, and the yield is $1.6\sigma$
compatible with the yield calculated in lattice QCD.
\item The decay analysis of the scalar isoscalar mesons shows that the assignment
of mesons to mainly-octet and mainly-singlet states is correct. Even
the production of mainly-octet scalar mesons - which should be
forbidden in radiative $J/\psi$ decays - peaks at 1865 MeV. The
decay analysis requires a small glueball content in the flavor wave
function of several scalar resonances. The glueball content as a
function of the mass shows a peak compatible with the peak in the
yield of scalar isoscalar mesons. The sum of the fractional glueball
contributions is compatible with one.
\item In the reaction $B_s\to J/\psi + K^+K^-$ reported by the LHCb collaboration, a primary $s\bar s$ couples to
mesons having a strong coupling to $K^+K^-$. Two peaks in the
$K^+K^-$ mass spectrum are seen due to $\phi(1020)$ and
$f_2'(1525)$,  but there is little evidence for the $f_0(1710)$ or
other high-mass scalar mesons coupling strongly to $K\bar K$.
High-mass scalar mesons are strongly produced by two initial-state
gluons but not by an $s\bar s$ pair in the initial state. They must
have sizable glueball fractions!
\item
The $D$ wave  amplitude in the  BESIII data on radiative $J/\psi$
decays reveales a high-mass structure which can be described by a
single Breit-Wigner or by the sum of three $\phi\phi$ resonances
interpreted as tensor glueballs a long time ago. The structure - and
further tensor resonances observed in radiative $J/\psi$ decays -
are tentatively interpreted as tensor glueball.
\item
In $J/\psi$ decays into $\gamma\pi^0\pi^0\eta'$ several resonances
are reported. The possibility is discussed that the pseudoscalar
glueball might be hidden in these data.\vspace{-6mm}
\end{itemize}
}
\maketitle
\section{Introduction}
\label{intro} The self-interaction between gluons leads to the
prediction of glueballs. Their masses are calculated by
discretization of QCD on a lattice or in approximate solutions of
QCD. A large number of glueball is predicted with the scalar
glueball as ground state, followed in mass by a tensor and a
pseudoscalar glueball. Typical masses of the lowest-mass glueballs
are shown
below:\\

\renewcommand{\arraystretch}{1.4}
\begin{tabular}{llllll}
\hline\hline

$0^{++}$ & 1710\er50\er\ 80\,MeV \cite{Morningstar:1999rf} &
1850\er130 MeV \cite{Huber:2021yfy}& 1980\,MeV
\cite{Szczepaniak:2003mr}& 1920\,MeV \cite{Rinaldi:2021dxh}\\
$2^{++}$ & 2390\er30\er120\,MeV  \cite{Morningstar:1999rf}&
2610\er180 MeV\cite{Huber:2021yfy} & 2420\,MeV
\cite{Szczepaniak:2003mr} & 2371\,MeV \cite{Rinaldi:2021dxh} \\
$0^{-+}$ & 2560\er35\er120\,MeV  \cite{Morningstar:1999rf}&
2580\er180 MeV \cite{Huber:2021yfy}
&2220\,MeV \cite{Szczepaniak:2003mr}&           \\
\hline\hline\\
\end{tabular}

The widths are essentially undetermined. The yields of glueballs
have been calculated for the process $J/\psi\to\gamma G$. The yields
are surprisingly high: the largest radiative yield of a meson,
$J/\psi\to\gamma \eta'$, is (5.25\er0.07)$\cdot 10^{-3}$. The yield
of the tensor glueball is expected to be
larger by a factor 2! The pseudoscalar yield is given for two glueball masses.\\

\renewcommand{\arraystretch}{1.4}
\begin{tabular}{lccccr}
\hline\hline
$BR_{J/\psi\to\gamma G_{0^{++}}}$(TH)&=& $(3.8\pm0.9)$&$\cdot 10^{-3}$&&\cite{Gui:2012gx}\\
                                                         &$\approx$& $3$&$ \cdot 10^{-3}$&&\cite{Narison:1996fm}\\
$BR_{J/\psi\to\gamma G_{2^{++}}}(TH)$&=& ($11\pm2)$&$\cdot 10^{-3}$&&\cite{Chen:2014iua}\\
$BR_{J/\psi\to\gamma G_{0^{-+}}}(TH)$&=& $(0.231\pm0.080)$&$\cdot 10^{-3}$& M=2395\,MeV&\cite{Gui:2019dtm}\\
 &=&$(0.107\pm 0.037)$&$\cdot 10^{-3}$&M=2560\,MeV&\cite{Gui:2019dtm}\\
\hline\hline\\
\end{tabular}

The scalar glueball is often supposed to intrude the spectrum of
scalar mesons and to mix with them. So far, authors have assumed
that the three scalar mesons $f_0(1370)$, $f_0(1500)$, $f_0(1710)$
contain the $1^3P_0$ $n\bar n$ and $s\bar s$ scalar mesons and the
glueball of lowest mass. Since only two scalar isoscalar mesons in
one nonet are expected, one of the three states seems to be
supernumerous, hence a glueball must have entered. The decay modes
of all three mesons are incompatible with one beeing a pure
glueball, hence these three mesons are supposed to mix \be
\begin{array}{r}
f_0(1370)  \\
f_0(1500)\\
f_0(1710) \\
\end{array}
= \left( \begin{array}{rrr}
x_{11} & x_{12} & x_{13} \\
x_{21} & x_{22} & x_{23} \\
x_{31} & x_{32} & x_{33} \\
\end{array}
\right)
\begin{array}{ll}
|n\bar n\rangle  \\
|s\bar s\rangle \\
|gg\rangle \\
\end{array}
\label{eq:mixEK} \ee where the mixing has to be determined from the
decay (and the production) of the three mesons.

\begin{figure*}[ph]
\includegraphics[width=\textwidth,height=0.25\textheight]{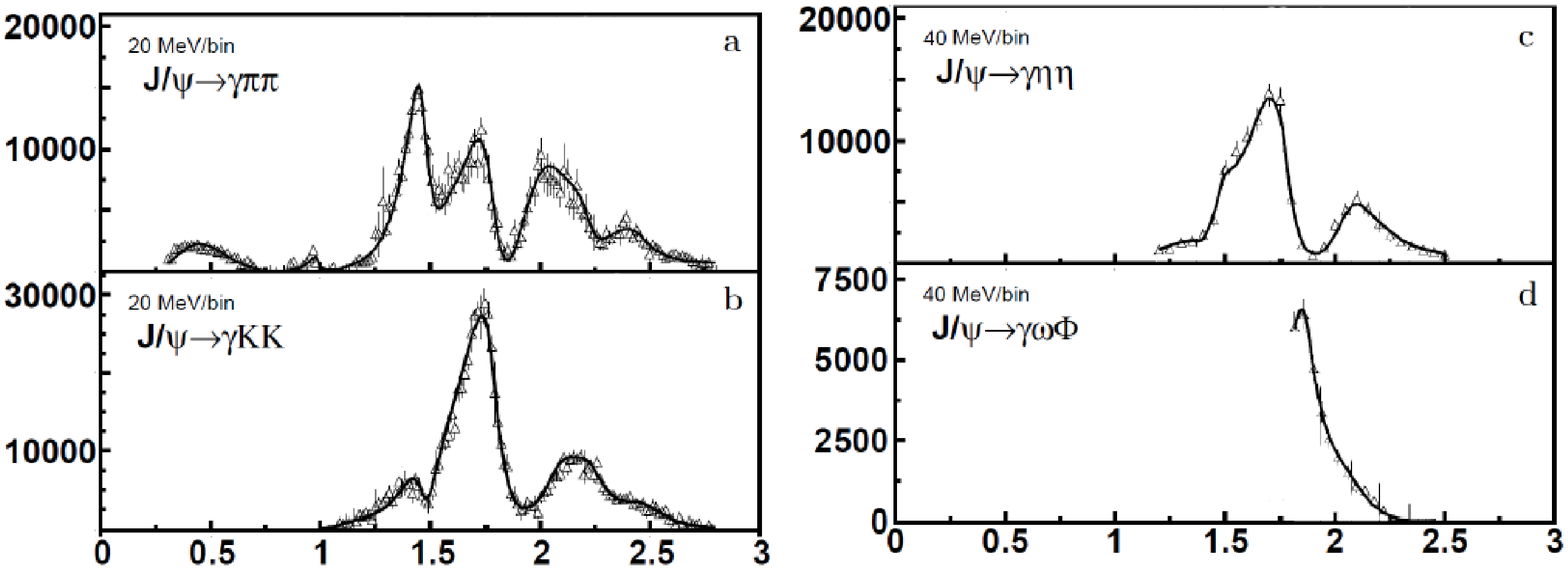}
\caption{The scalar intensity in $J/\psi\to \gamma\pi^0\pi^0$ (a),
$K^0_sK^0_s$ (b), $\eta\eta$ (c) and to $\omega\phi$ (d) from
Ref.~\cite{BESIII:2015rug,BESIII:2018ubj}. The  curves represent the
best fit \cite{Sarantsev:2021ein}.}
\label{fig-1}       
\centering
\includegraphics[width=0.95\textwidth,height=0.55\textheight]{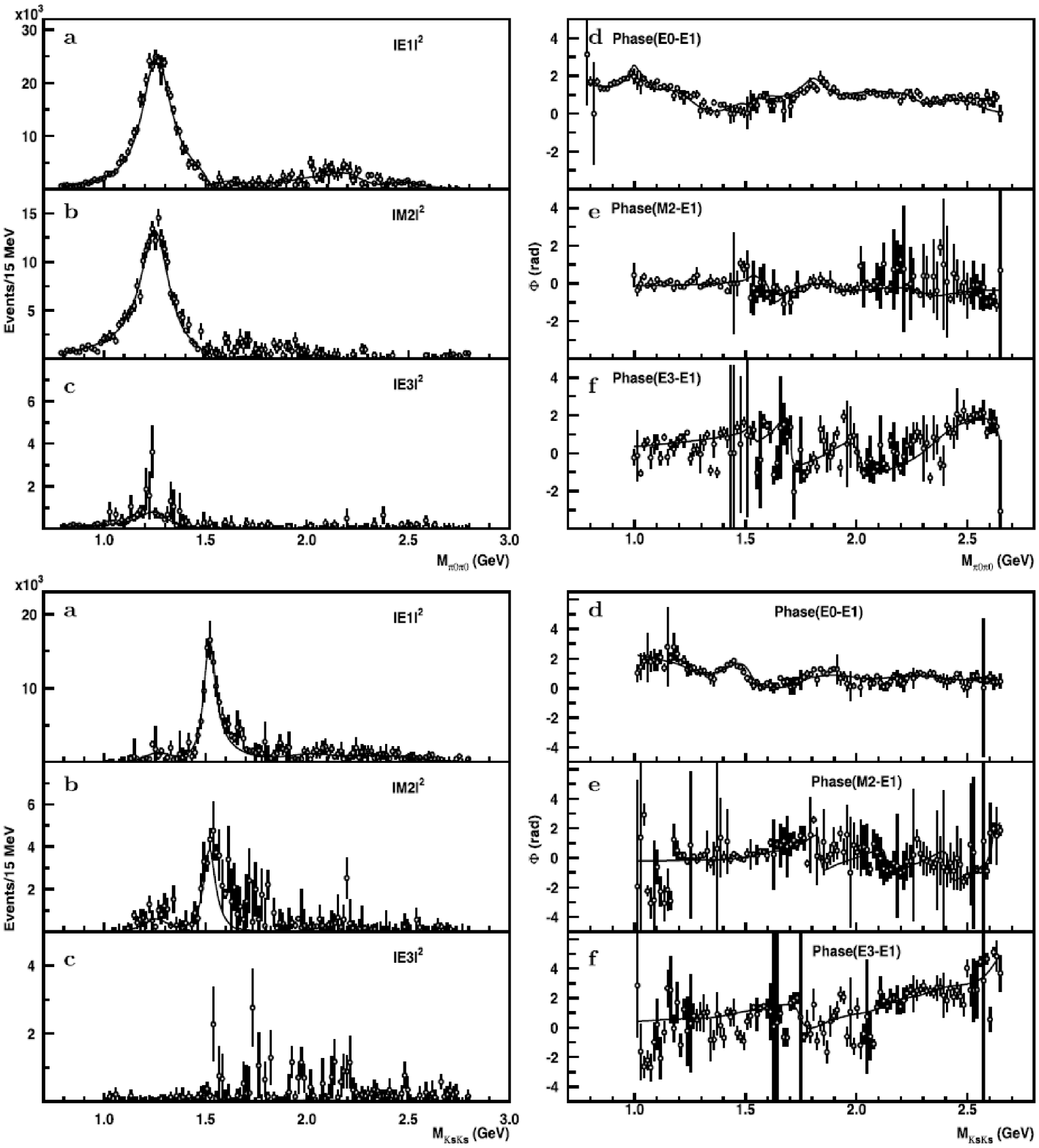}
\vspace{-5mm} \caption{\label{jpsia}$D$-wave intensities and phases
for radiative $J/\psi$ decays into $\pi^0\pi^0$ (top subfigures) and
$K_s\,K_s$ (bottom subfigures)  from
Ref.~\cite{BESIII:2015rug,BESIII:2018ubj}. The subfigures show the
$E1$ (a), $M2$ (b) and $E3$ (c) squared amplitudes and the phase
differences between the $E0$ and $E1$ (d) amplitudes, the $M2$ and
$E1$ (e) amplitudes, and the $E3$ and $E1$ (f) amplitudes as
functions of the meson-meson invariant mass. The phase of the $E0$
amplitude is set to zero. The  curves represent the best fit
\cite{Klempt:2022qjf}. }
\end{figure*}

\section{Data and coupled channel analysis} \label{sec-1} Radiative
decays of $J/\psi$ mesons is the prime source to discover glueballs.
$J/\psi$ mesons can convert into three gluons or into one photon and
two gluons. These two gluons interact and should form glueballs --
if glueballs exist. In Ref.~\cite{Sarantsev:2021ein} a
coupled-channel analysis of a large number of data sets was
reported. Of particular importance are the results from an amplitude
analysis of $J/\psi\to \gamma\pi^0\pi^0$~\cite{BESIII:2015rug} and
$K^0_sK^0_s$~\cite{BESIII:2018ubj}, supported by BESIII data on
$J/\psi\to\eta\eta$ and to $\omega\phi$. The CERN-Munich data on
$\pi\pi$ elastic scattering with the $\pi\pi$ $S$-wave are mandatory
to get the right solution. The GAMS data, in particular those on
$\pi^0\pi^0$ production from high-energy $\pi N$ scattering force
the solution above 1.6\,GeV. The data on $K_{\rm e4}$ decays fix the
low-mass part of the amplitude. Furthermore, 15 Dalitzplot from
$p\bar p$ annihilation at rest are included in the analysis as well
as BNL data on $\pi\pi\to K^0_sK^0_s$ and further GAMS data.
References to the data can be found in
Ref.~\cite{Sarantsev:2021ein}.

Figure~\ref{fig-1} shows invariant mass distributions of the BESIII
data assigned to the $S$-wave, Figure \ref{jpsia} (left) the
$D$-wave for $\pi^0\pi^0$ and $K^0_sK^0_s$.  The $D$ wave can be
excited by three electromagnetic amplitudes, $E1, M2$, and $E3$,
where the $E1$ amplitude is the most significant one. These three
amplitudes and relative phases are shown in Fig.~\ref{jpsia}
(right).

\section{The scalar glueball} \label{sec-2}
\begin{figure*}[ph]
\sidecaption
\includegraphics[width=0.6\textwidth,height=0.38\textwidth]{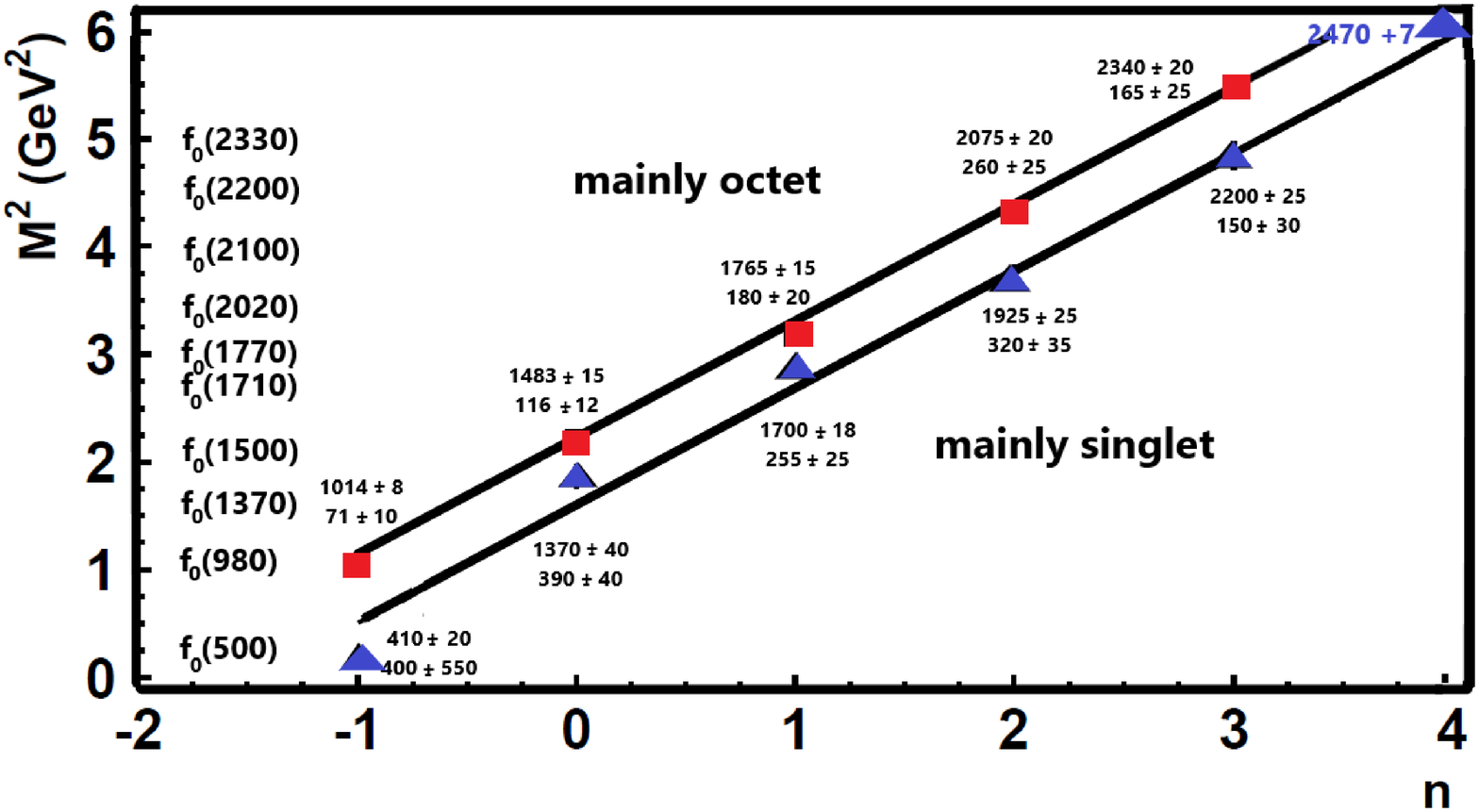}
\caption{The square masses of scalar mesons as a function of a
consecutive number~\cite{Sarantsev:2021ein}. The numbering starts
with -1 since $f_0(500)$ and $f_0(980)$ may not belong to the
series. Masses and widths are given in small numbers. The SU3
assignment is discussed in the text. The meson at 2470\,MeV was seen
by BESIII in $J/\psi\to\gamma \eta'\eta'$ decays
\cite{BESIII:2022zel}. Due to its decay it likely belongs to the
mainly singlet states. }
\label{fig-3}       
\end{figure*}

From the fit, existence and properties are scalar mesons are
deduced. Figure~\ref{fig-3} shows their squared masses as a function
of a consecutive number. The mesons are grouped pairwise. The
higher-mass states are interpreted as octet states in SU3, the
lower-mass states as singlet states. These assignments are verified
in an analysis of their decays discussed below. The scalar glueball
is identified by three different methods.

\begin{figure*}[pt]
\begin{tabular}{ccc}
\hspace{-2mm}\includegraphics[width=0.33\textwidth,height=0.2\textwidth]{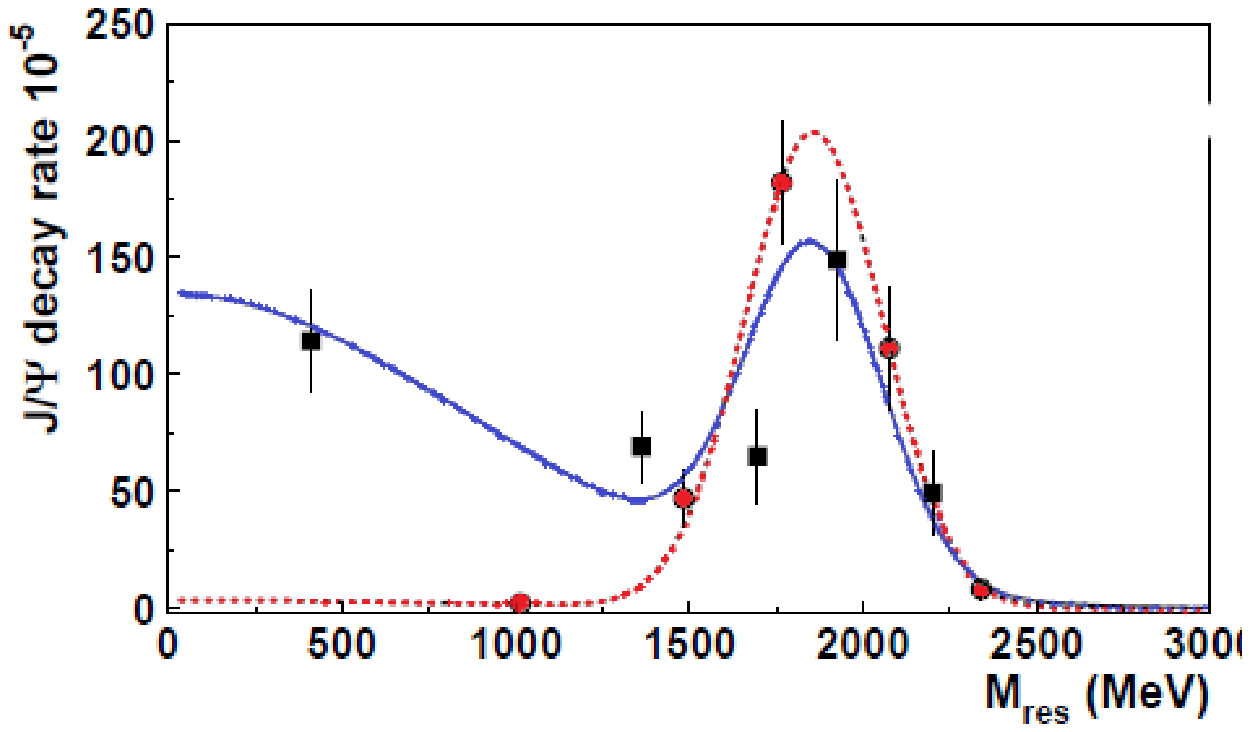}&
\hspace{-4mm}\includegraphics[width=0.33\textwidth,height=0.2\textwidth]{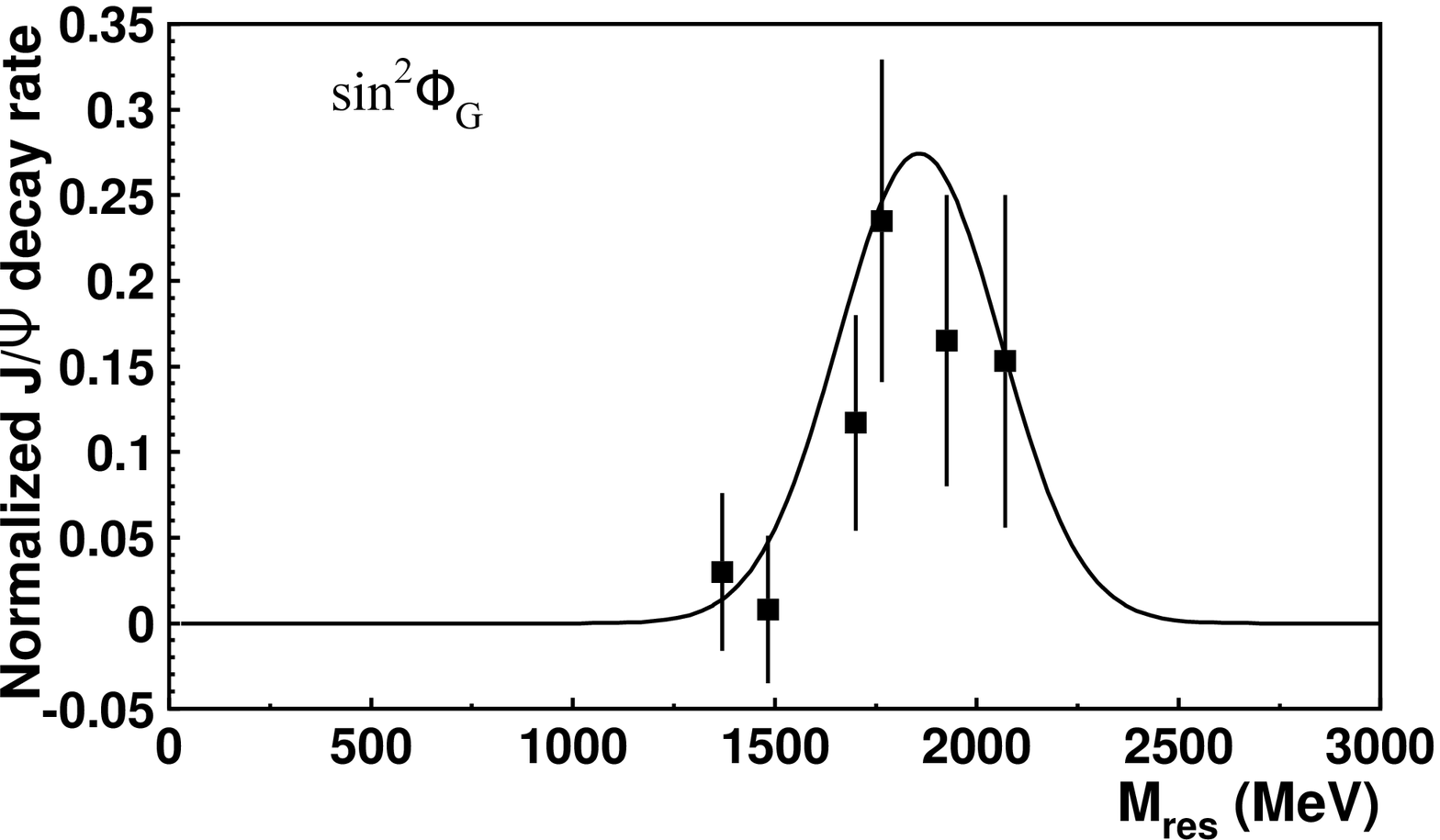}&
\hspace{-4mm}\includegraphics[width=0.33\textwidth,height=0.2\textwidth]{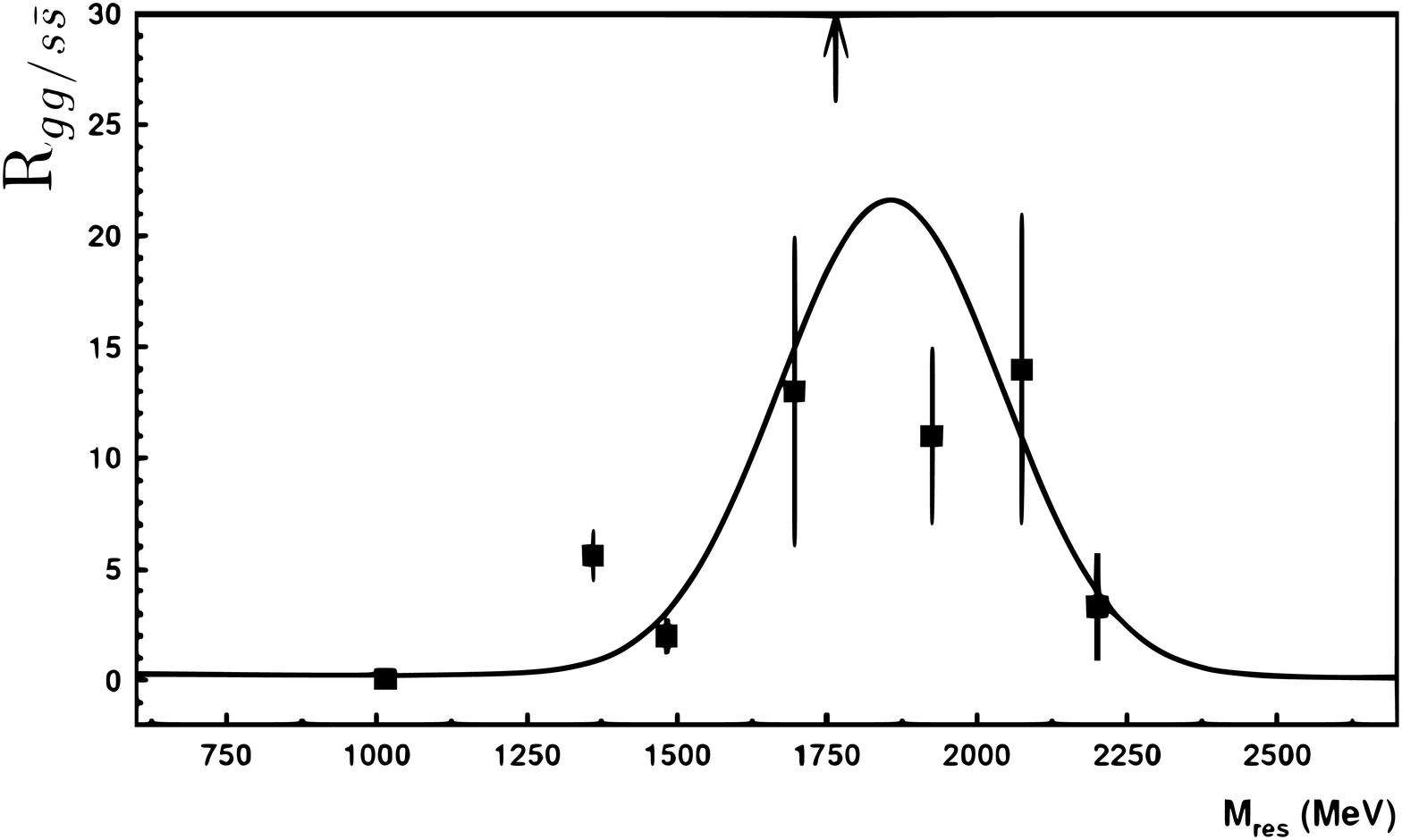}\vspace{-2mm}
\end{tabular}
\caption{The scalar glueball is identified by three methods: the
radiative yield of $J/\psi$ decays into scalar singlet and isoscalar
octet mesons peaks at 1.865\,MeV (left); the glueball component in
the mesonic wave function of scalar mesons peaks at the same mass
(center, \cite{Klempt:2021wpg}); the ratio of the yields of scalar
mesons in radiative $J/\psi$ decays and in $B_s\to J/\psi\ +$ scalar
mesons peaks as well (right).\vspace{-2mm}}
\label{fig-4}       
\end{figure*}

\paragraph{The scalar glueball from the yield in radiative $J/\psi$ decays:}
First, the total radiative yield of scalar mesons is plotted as a
function of their mass for both mainly-octet and mainly-singlet
mesons (see Fig.~\ref{fig-4}, left). Radiative production of
mainly-octet mesons should be suppressed: two gluons cannot convert
into an octet meson. The fact that mainly-octet mesons are produced
at the same rate as mainly-singlet mesons can only be explained when
these mesons are dominantly produced via their glueball component.
The total yield is derived from the fit described here and from
known radiative decay branching ratios of the $J/\psi$ meson.

\begin{figure*}[ph]
\sidecaption
\includegraphics[width=0.6\textwidth,clip]{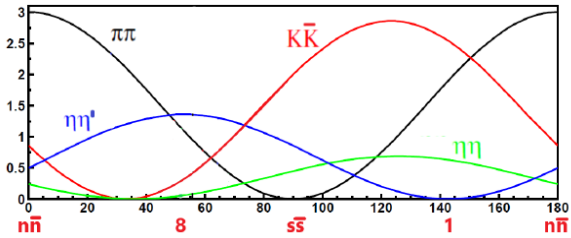}
\caption{Squared coupling constants $\gamma_\alpha$ for meson decays
into a pair $\alpha$ of pseudoscalar mesons \cite{Klempt:2021wpg}.
For the mixing angle 90$^\circ$, e.g., the meson is a pure $s\bar s$
state and there are no $\pi\pi$ decays.}
\label{fig-5}       
\end{figure*}

A fit to the yield shown in Fig.~\ref{fig-4} (left) with a
Breit-Wigner function returns mass, width and yield: \be
M-\frac12\Gamma =(1865\pm 25^{\,+10}_{\,-30})-i (185\pm
25^{\,+15}_{\,-10}){\rm \,MeV},\quad Y = (5.8\pm 1.0)\,10^{-3}. \ee
\paragraph{The scalar glueball from the decays of scalar mesons:}
The second method exploits the decay modes of the scalar
mesons~\cite{Klempt:2021wpg}. The decay modes depend on the mixing
angle of the singlet and the octet isoscalar mesons belonging to the
same nonet. Figure~\ref{fig-5} shows the squared coupling constants
for meson decays into two pseudoscalar mesons as a function of the
mixing angle.

The wave function of a meson can be decomposed into its $n\bar n$,
$s\bar s$ and glueball component $G$, where $\varphi^{\rm s} _{\rm
n}$ is the scalar mixing angle in nonet $n$, $\phi^G _{\rm nL}$ and
$\phi^G _{\rm nH}$ are the meson-glueball mixing angles of the
low-mass state L and of the high-mass state H in the nth nonet.
$\sin^2\phi^G _{\rm nH}$ and $\sin^2\phi^G _{\rm nL}$ are the
glueball contents of the two mesons. \beq f^{\rm nH}
_0(xxx)=\left(n\bar n\cos\varphi^{\rm s} _{\rm n}-s\bar
s\sin\varphi^{\rm s} _{\rm n}\right )\cos\phi^G _{\rm nH} +
G\sin\phi^G _{\rm nH}\nonumber\eeq \beq f^{\rm nL}
_0(xxx)=\left(n\bar n\sin\varphi^{\rm s} _{\rm n}+s\bar
s\cos\varphi^{\rm s} _{\rm n}\right )\cos\phi^G _{\rm nL} +
G\sin\phi^G _{\rm nL}\nonumber\eeq

The coupling of a meson to the final state $\alpha$ can be written
as $g_\alpha ^{\rm n} = c_{\rm n} \gamma_\alpha ^q +
c_G\gamma_\alpha ^G$. Here, the $q\bar q$ and the glueball
components of a scalar meson couple with the SU(3) structure
constants $\gamma_\alpha$ and with a decay coupling constant $c_n$
or $c_G$ to the final states $\alpha$. A fit to the decay branching
ratios yields the glueball fractions of the scalar
mesons~\cite{Klempt:2021wpg}. These are shown in Fig.~\ref{fig-3}.
The glueball fractions derived from the decay of scalar mesons into
two pseudoscalar mesons shows a peak with a shape that is fully
compatible with the peak in the yield of scalar mesons in radiative
$J/\psi$ decays. In $f_0(1370)$-$f_0(1500)$-$f_0(1710)$ mixing
scenarios, it is imposed that the full glueball is distributed over
the three mesons. Here, the sum of the glueball fractions is
determined to 0.78\er 0.18. The information of the highest-mass
scalar meson is insufficient to determine a glueball content. Since
this part is missing, nearly the full glueball is covered and
distributed over the mesons observed in radiative $J/\psi$ decays.

\paragraph{The scalar glueball from a comparison with $B_s\to J/\psi f_0$ and
$J/\psi\to \gamma f_0$:}

\begin{figure*}[ph]
\includegraphics[width=0.9\textwidth,clip]{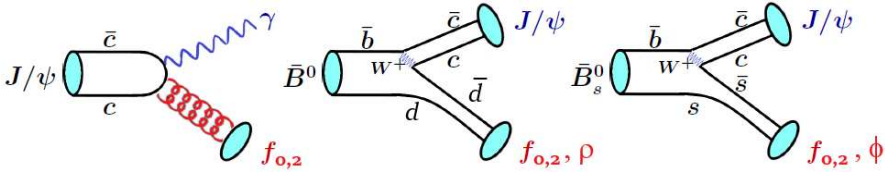}
\caption{Scalar meson production in $B^0\to J/\psi\,f_0$, $B_s\to
J/\psi\,f_0$, $J/\psi\to \gamma f_0$.\vspace{-3mm}}
\label{fig-6}       
\end{figure*}

\begin{figure*}[pt]\centering
\includegraphics[width=0.85\textwidth,clip]{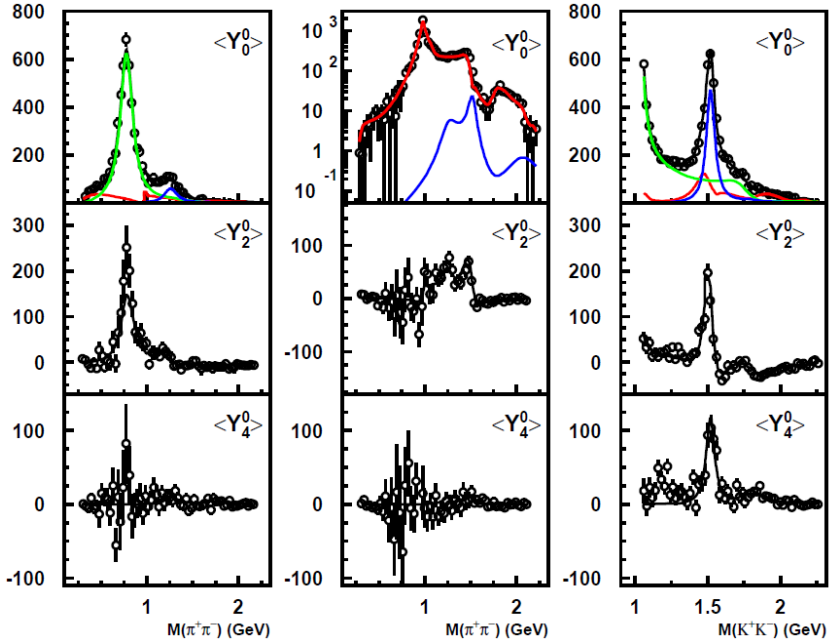}
\caption{(Color online) The spherical harmonic moments $\cos
\theta_{h\bar h}$ for $\bar B^0\leftrightarrow B^0\to J/\psi
\pi^+\pi^-$ (left, \cite{LHCb:2014vbo}), $\bar B^0_{s}
\leftrightarrow B^0_{s}\to J/\psi \pi^+\pi^-$ (center,
\cite{LHCb:2014ooi}), and into $J/\psi\,K^+K^-$ (right,
\cite{LHCb:2017hbp}). The points with error bars represent the LHCb
data, the solid curve  our fit. Partial waves with $L\geq 6 $ and
Odd partial waves are not included in the fit. The $S$-wave
($P$-wave, $D$-wave) contribution is shown as red (green, blue)
curve.}
\label{fig-7}       
\end{figure*}

In $J/\psi$ radiative decays, a photon is emitted and two gluons are
created which convert into an observable meson. In $B^0$ and $B_s$
decays into a $J/\psi$ meson recoiling against a $\pi^+\pi^-$ or
$K^+K^-$ pair, a primary $d\bar d$ or $s\bar s$ pair is produced
which form a meson (see Fig.~\ref{fig-6}).
The LHCb collaboration
studied this process and the mixing of $B^0$ and $B_s$ mesons with
their antiparticles and determined the decay-time-dependent CP
asymmetry. The collaboration also presented spherical harmonic
moments and their dependence of the $\pi^+\pi^-$ or $K^+K^-$
invariant mass and interpreted these results in terms of
contributing resonances. These data were included in the
coupled-channel analysis of Ref.~\cite{Sarantsev:2021ein}. The data
and the fit are shown in Fig.~\ref{fig-7}.

The scalar intensity is surprisingly small, in particular above
1600\,MeV. The $f_0(1710)/f_0(1770)$ complex that is seen so
strongly in $J/\psi\to\gamma f_0(1710)/f_0(1770)\to\gamma K\bar K$
is nearly absent when a primary $s\bar s$ pair is present at the
same invariant mass and under similar kinematical conditions. The
wave function of high-mass mesons obviously have little overlap with
a local $s\bar s$ pair. However, a small glueball fraction in a
mesonic wave function enhances the chance to produce this meson in
radiative $J/\psi$ decays.

In Fig.~\ref{jpsia} (right), the ratio of frequencies for scalar
meson production in $J/\psi\to \gamma f_0$ and $B_s\to J/\psi\,f_0$
decays is shown as a function of the scalar meson mass. The
distribution is compared to the Breit-Wigner shape determined from
the yield distribution in Fig.~\ref{fig-4}. The ratio is fully
compatible with the Breit-Wigner fit, even though with limited
statical accuracy.

\paragraph{The scalar glueball summary:}
The scalar glueball is identified from the yield of scalar mesons in
radiative $J/\psi$ decays, from the decay-mode analysis of scalar
mesons, and from a comparison of the yield of scalar mesons in
$J/\psi\to \gamma f_0$ and $B_s\to J/\psi\,f_0$ decays.

\section{Evidence for the tensor glueball} \label{sec-3}

Figure~\ref{jpsia} shows large signals due to $f_2(1270)\to\pi\pi$
and $f_2'(1525)\to K\bar K$. In the $\pi\pi$ invariant mass spectrum
a further enhancement is seen with mass, width and yield \be
M-\frac12\Gamma = (2210\pm 60) -i (180\pm 60) {\rm \,MeV},\quad Y =
(0.35\pm 1.0)\,10^{-3}, \ee
where the yield is the sum of the $\pi\pi$ and $K\bar K$
contributions. Again, this peak is absent in $B^0$ and $B_s$ decays.
Hence it is likely produced due to a small tensor-glueball component
in the wave function. The yield is low even if all intensity
reported from $J/\psi\to \gamma f_2$ decays above 1.8\,GeV is added
\cite{Klempt:2022qjf}. The peak can be fitted as sum of the three
tensor resonances reported in the doubly OZI rule violating process
$\pi^-p\to \phi\phi n$. The three states were proposed as one, two
or three tensor glueballs \cite{Etkin:1987rj}.

\section{How to find the pseudoscalar glueball } \label{sec-4}
\begin{figure*}[h!]
\sidecaption
\includegraphics[width=0.65\textwidth,height=0.45\textwidth]{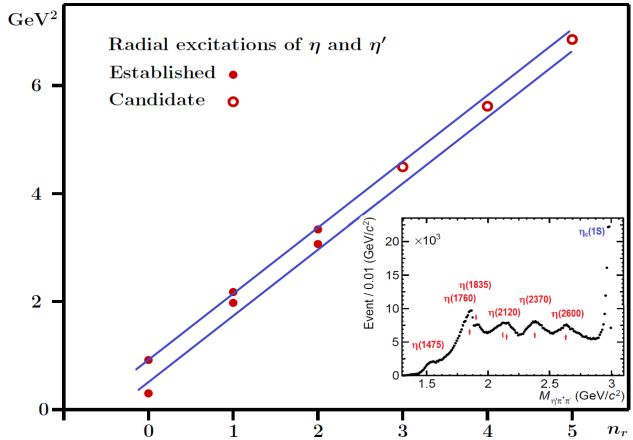}
\caption{Left: Suggested $M^2$-versus-n plot of structures
interpreted as pseudoscalar mesons. The data from
Ref.~\cite{BESIIICollaboration:2022kwh}) are shown as inset.}
\label{fig-8}
\vspace{-5mm}     
\end{figure*}

In the $\pi^+\pi^-\eta^\prime$ invariant mass distribution observed
in $J/\psi\to \pi^+\pi^-\eta^\prime$, a series peaks is
observed~\cite{BESIIICollaboration:2022kwh} (see inset of
Fig.~\ref{fig-8}). We tentatively assume that they have pseudoscalar
quantum numbers. Two linear trajectories can be drawn. The
resonances above 2\,GeV might be split: the $\eta'f_0(1500)$ isobar
would be an octet-like state, $\eta'f_0(1370)$ a singlet-like state.
The separation would require a cut in the $\pi\pi$ invariant mass
into a region below and above 1480\,MeV. Then, peaks at slightly
different masses should be seen in the $\pi^+\pi^-\eta^\prime$
invariant mass distribution. High-mass mesons of low spins are often
suppressed in hadronic reactions. If they have a glueball component,
their production in radiative $J/\psi$ decays could be significant.
Hence I suggest that the pseudoscalar glueball may be wide and that
at least all peaks above 2\,GeV in Fig.~\ref{fig-8}a have a
significant glueball fraction in their wave function.

\section{Conclusions} \label{sec-5}
Radiative $J/\psi$ decay have revealed the scalar glueball and show
evidence for the tensor glueball. First traces of the pseudoscalar
glueball may have shown up as well. The high statistics now
available from BESIII and modern coupled-channel analyses have the
chance to observed more decay modes and to improve our understanding
of the glueball
spectrum.\\[3ex]

{\it I would like to thank the organizer of this conference for the
kind invitation to this interesting conference in such a beautiful
place.}

\bibliography{New-references}

\end{document}